\documentclass[aps, prb, superscriptaddress, twocolumn, longbibliography, showpacs, preprintnumbers,
amssymb, floatfix]{revtex4-1}

\usepackage{graphicx}
\usepackage{epsfig}
\usepackage{amsmath}
\usepackage{amssymb}
\usepackage{amsbsy}
\usepackage{xcolor}
\usepackage{soul}
\usepackage{hyperref}

\newcommand{\spvec}[1]{\ensuremath{\mathbf{#1}}}
\newcommand{\unitvec}[1]{\ensuremath{\mathbf{\hat{#1}}}}

\newcommand{\colvec}[1]{\ensuremath{\mathrm{#1}}}

\newcommand{\commentout}[1]{{}}

\newcommand{\beq}{\begin{equation}}
\newcommand{\eeq}{\end{equation}}

\newcommand{\cbE}{\boldsymbol{\mathbf{\cal E}}}
\newcommand{\cbH}{\boldsymbol{\mathbf{\cal H}}}

\begin{document}

\title{Strong radiative interactions and subradiance in disordered metamaterials}
  \date{\today }

\author{Stewart D. Jenkins}
\affiliation{Mathematical Sciences and Centre for Photonic Metamaterials, University of Southampton, Southampton SO17 1BJ, United Kingdom}
\author{Nikitas Papasimakis}
\affiliation{Optoelectronics Research Centre and Centre for Photonic Metamaterials, University of Southampton, Southampton SO17 1BJ, United Kingdom}
\author{Salvatore Savo}
\affiliation{Optoelectronics Research Centre and Centre for Photonic Metamaterials, University of Southampton, Southampton SO17 1BJ, United Kingdom}
\affiliation{TetraScience Inc., 114 Western Ave, Boston, MA, 02134}
\author{Nikolay I. Zheludev}
\affiliation{Optoelectronics Research Centre and Centre for Photonic Metamaterials, University of Southampton, Southampton SO17 1BJ, United Kingdom}
\affiliation{Centre for Disruptive Photonic Technologies, School of Physical and Mathematical Sciences and The Photonics Institute,
            Nanyang Technological University, Singapore 637378, Singapore}
\author{Janne Ruostekoski}
\affiliation{Mathematical Sciences and Centre for Photonic Metamaterials, University of Southampton, Southampton SO17 1BJ, United Kingdom}
\affiliation{Department of Physics, Lancaster University, Lancaster, LA1 4YB, United Kingdom}

\begin{abstract}
We provide detailed comparisons between experimental findings and numerical simulations of large cooperatively interacting, spatially disordered metamaterial arrays, consisting of asymmetrically split rings. Simulation methods fully incorporate strong field-mediated inter-meta-atom interactions between discrete resonators and statistical properties of disorder, while approximating the resonators' internal structure. Despite the large system size, we find a qualitative agreement between the simulations and experiments, and characterize the microscopic origins of the observed disorder response. Our microscopic description of macroscopic electrodynamics reveals how the response of disordered arrays with strong field-mediated interactions is inherently linked to their cooperative response to electromagnetic waves where the multiple scattering induces strong correlations between the excitations of individual resonators.
Whereas for a regular array the response can be overwhelmingly dominated by a spatially-extended collective eigenmode with subradiant characteristics, a gradual increase of the positional disorder rapidly leads to a spatial localization of both the electric and magnetic dipolar excitation profile of this eigenmode. We show how the effects of disorder and cooperative interactions are mapped onto the transmission resonance in the far field spectrum and measure the ``cooperative Lamb shift'' of the resonance that is shifting toward the red as the disorder increases. The interplay between the disorder and interactions generally is most dramatic in the microwave arrays, but we find that in suitable regimes the strong disorder effects can be achieved also for plasmonic optical systems.
\end{abstract}
\maketitle

\section{Introduction}

According to conventional wisdom, disorder and interactions are undesired phenomena with deleterious effects on the design and development of electromagnetic (EM) devices and functionalities. They lead to uncontrolled coupling between radiation and matter, resonance broadening, shifts and dephasing, thus limiting the potential of devices for practical applications, such as sensing and telecommunications. Intrinsic disorder in optical materials affects the transport properties of light and can even lead to the absence of diffusive wave propagation~\cite{andersonlight}, analogously to the Anderson localization of electrons in solids. In artificial materials, in particular, disorder is  known to introduce unwanted scattering and deteriorate the performance of optical devices; however, at the same time, engineered EM materials provide unique controlled environments for studying the effects of disorder and implementing novel disorder-induced functionalities \cite{Wiersma2013}.

The advent of metamaterials allows one to manipulate the EM response across different scales. At the microscopic level of individual unit-cell resonators (meta-molecules), electric and magnetic multipolar properties can be engineered. In large metamaterial arrays the collective behavior of the ensemble of resonators can be altered due to strong radiation-mediated interactions. Systems where the EM-field mediated interactions are not weak have been utilized in regular metamaterial arrays, e.g., in subdiffraction focusing~\cite{Sentenac2008,KAO10}, metalensing~\cite{lemoult2010}, generation of coherent, collimated beams~\cite{ZheludevEtAlNatPhot2008,AdamoEtAlPRL2012}, in narrow transmission resonances~\cite{FedotovEtAlPRL2010,PapasimakisEtAlAPL2009,JenkinsLineWidthNJP,CAIT,ValentineNatComm2014,wang_nanoshells}, in subradiance of few-resonator systems~\cite{Lovera, FanCapasso, Frimmer, Watson2016} and of massive spatially-extended samples~\cite{Jenkins2017PRL}, atomic lattices~\cite{Jenkins2012a,Facchinetti}, in superconducting quantum-interference devices~\cite{Anlageprx,Anlagepre17,Ustinov2015,Lazarides2013}, and in thin semiconductor layers~\cite{Imamoglu2018,Imamoglu2018b}.
Introducing disorder in metamaterials typically concerns either stochastic distributions of the resonance frequencies (inhomogeneous broadening)~\cite{PhysRevE.73.056605,Zharov2005,Gollub,JenkinsRuostekoskiPRB2012b,Ustinov2015,Fistul2017} or positional disorder~\cite{Aydin2004,papasimakis2009,Helgert2009,Singh2010,Engheta2010, Albooyeh2010,SavoEtAlPRB2012,Zhang2017,Marini2016,Albooyeh2014,Tretyakov16,Jang2018,Pinheiro2017}. Positional disorder fundamentally differs from the inhomogeneous broadening in that the latter can only reduce the role of light-mediated interactions between the resonators~\cite{JenkinsRuostekoskiPRB2012b}, while positional disorder can dramatically change their collective nature.
However, most experiments -- as well as applications -- of disorder have not exploited strong field-mediated interactions in metamaterial systems, while
theoretical analysis of positionally disordered metamaterials has typically focussed on the effects of disorder on the metamaterial effective parameters.
The description of metamaterials with effective parameters based on a continuous medium approach treats the interactions in an average sense, where the
precise information of the locations of the discrete resonators is lost (in violation of the fact
that the resonant dipole-dipole interactions between the resonators sensitively depend on their spatial separation).
While such approaches often work for weakly interacting resonator systems, in strongly interacting systems the response can substantially differ~\cite{Javanainen2014a,JavanainenMFT}.

Here we apply large-scale numerical simulations for the microscopic description of the electrodynamics of cooperatively responding disordered metamaterial arrays and compare the simulations with far-field experimental measurements.
Our work on a large disordered array of asymmetrically split rings (ASRs) is motivated by previous experimental observations of the difference in between the far- and near-field responses of ordinary `incoherent' metamaterials and interacting `coherent' metamaterials~\cite{papasimakis2009,SavoEtAlPRB2012}.
We show how their response is inherently linked to strong field-mediated interactions between the resonators that induce correlations between the excitations of different resonators and cannot be described by effective continuous medium theories of electrodynamics.
We analyze the system by using simulation methods that treat each discrete meta-atom individually and fully incorporate all recurrent scattering events between all the resonators as well as the statistical properties of the positional disorder, while approximating the resonators' internal structure.
The approach allows us to characterize the microscopic principles of the collective macroscopic EM response of the disordered metamaterial.
We show how the interplay between the strong interactions and the disorder leads to emergent behavior and correlated response that is qualitatively distinct from the response of the individual meta-molecules and that of regular arrays.

Due to the strong collective response of the metamaterial, we find that the effects of disorder are much more dramatic than one would surmise based on the disciplined and predictable responses of regular arrays.
The dramatic difference between the responses of regular and disordered arrays manifests itself most clearly in the microscopic properties of collective excitation eigenmodes.
For instance, in a regular array, the transmission resonance corresponds to a single, giant, spatially-extended subradiant eigenmode~\cite{Jenkins2017PRL}.
Even small amounts of disorder lead to rapid spatial shrinking of this cooperative excitation eigenmode.
With increasing disorder this eigenmode continuously deforms from a uniform excitation to a strongly localized one that, nevertheless, is influenced by the resonators in the entire array.
The analysis shows how the scattered fields directly convey information about the positional disorder and interactions. We link the contrasts, widths, and shifts of the resonances to the microscopic properties of the collective excitation eigenmodes. For instance, the measured shift and its calculated disorder-dependent statistical fluctuations represent the detection of a ``cooperative Lamb shift'' \cite{Friedberg1973} in this system, which has
been actively measured in various ensembles of resonant emitters, e.g., of nuclei~\cite{ROH10}, ions~\cite{MeirLamb}, thermal atoms in vapor cells~\cite{Keaveney2012}, and cold trapped atoms~\cite{Jennewein_trans,Roof16,Ye2016,Dalibard_slab}.

We have discovered that in the ASR arrays the strong collective interactions can set in surprisingly easily. The cooperative effects are not only restricted to microwave metamaterials consisting of metallic resonators, where the Ohmic losses are weak, but surprisingly, with appropriate engineering of the microscopic resonator properties, we can identify suitable parameter regimes of strong disorder effects and intense light confinement even in metallic materials in the optical regime (plasmonics).

\section{Collective interactions in disordered resonator arrays}

\begin{figure}
  \centering
  \includegraphics[width=0.55\columnwidth]{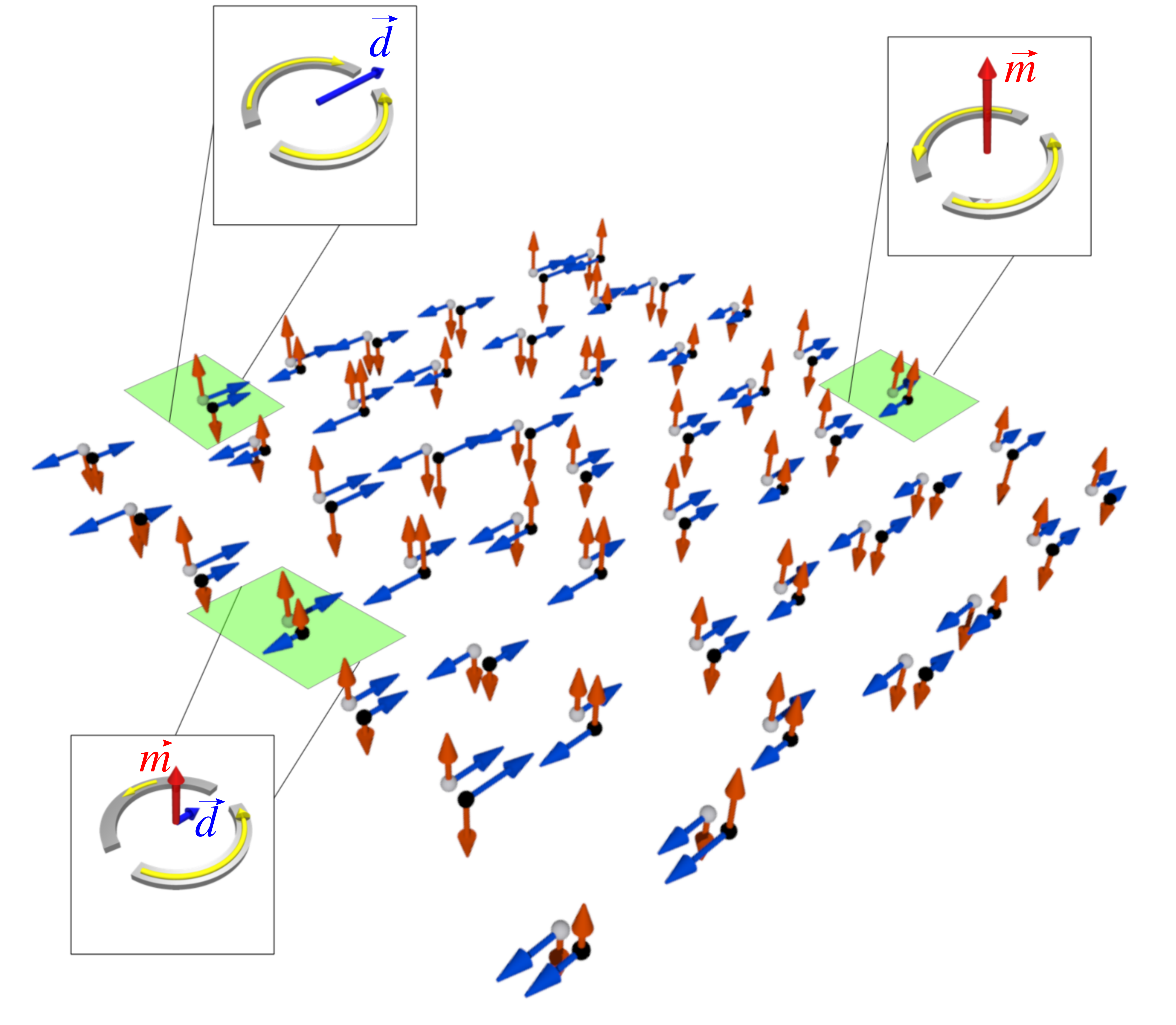}
  \caption{\textbf{Schematic diagram of the theoretical model.}
    A disordered array of meta-molecules, with each unit-cell consisting of
    two concentric arcs.  Currents flowing symmetrically (antisymmetrically)
    in a meta-molecule produce an oscillating electric (magnetic) dipole
    [upper-left (right) inset].  Each arc is governed by a
    single dynamic variable for the electric current whose oscillation generates both electric (blue arrows)
    and magnetic (red arrows) dipoles. The varying length of arrows across the array illustrates how
    disorder in meta-molecule positions can yield a non-uniform
    response to a uniform incident field.
  }
  \label{fig:schematic}
\end{figure}

\subsection{Ordered and disordered arrays}

Consider a planar metamaterial composed of plasmonic meta-molecules.
Generally, the response of each resonator to an applied
EM field has both electric and magnetic characteristics.
Plasmonic oscillations in each meta-molecule scatter EM fields which then
drive plasmonic oscillations in other resonators in the array.
Fields scattered between the resonators mediate
long-range interactions between them \cite{JenkinsLongPRB}.  Strong
interactions, particularly pronounced for
meta-molecules separated by a fraction of a wavelength, cause the
metamaterial to respond collectively to an
incident driving field.

In a regular array, the lattice structure imposes a
discrete translational symmetry on the interactions between
meta-molecules that comprise a unit-cell.
The resulting regularity in the interactions favours a collective
response to an incident plane wave in which all meta-molecules
oscillate with equal amplitudes and a spatially coherent phase.
Since the spatial distribution of meta-molecule
excitations remains uniform regardless of the strength of unit-cell
interactions, the lack of variation partially obscures the
role of collective interactions in the regular array.

In a disordered metamaterial, the interactions between meta-molecules no longer
lead to uniform response of the array.
Since these interactions strongly depend on
the meta-molecules' relative positions, introducing disorder to their positions breaks
the discrete symmetry of the interactions present in the regular array.
This variation in interaction strength means that the distribution of
meta-molecule excitations is no longer uniform, but can exhibit localized
excitations with intensities stronger than those attainable in a
regular array.

Collective excitations  due to EM-mediated interactions in a regular planar array of ASRs were identified in terms of a many-body subradiant eigenmode in Ref.~\onlinecite{Jenkins2017PRL}.
An incident field, normal to the array, was able to excite a spatially extended eigenmode comprising the entire lattice of over 2000 resonators, indicating a giant realization
of the suppressed emission, originally introduced by Dicke~\cite{Dicke54}.
This massive correlated radiative excitation also violates standard effective continuous medium descriptions of electrodynamics.
Although the translational
symmetry of the regular array provides a response displaying little obvious visible signs of collective effects, the existence of the giant many-body subradiant mode was manifested in the properties of the far-field transmission resonance. The collective response resulted from the interplay between the electric and magnetic dipole excitations in the ASRs: although the incident field directly only coupled to the spatially uniform electric dipole excitations, the intrinsic asymmetry in the arc lengths of the ASR enabled the transfer of the excitation to an eigenmode with a nearly uniform magnetic dipole excitation and suppressed radiative decay.

\subsection{Microwave and plasmonic materials}

In both regular and disordered arrays, long-range interactions develop
between excitations in different resonators due to the multiple scattering
of EM waves in the system.
Meta-molecules that suffer substantial radiative losses in isolation thus
strongly interact when placed in the array. Energy lost to non-radiative losses,
by contrast, inhibits interactions.
Non-radiative losses limit the number of times an EM wave can scatter
between resonators before it is absorbed into the material~\cite{JenkinsLongPRB}.

In microwave metamaterials, non-radiative losses mainly occur in the dielectric substrate
surrounding the meta-molecules rather than in the meta-molecules
themselves \cite{FedotovEtAlPRL2007}.
In the infrared and optical parts of the spectrum, by contrast,
dissipation severely limits quality factors of metallic
meta-molecule resonances.
Ohmic losses impose a lower bound on the quality factors of collective
plasmonic metamaterial resonances, generally limiting the scope of
their potential applications \cite{West2010}.
We show how this generally leads to significant suppression of the long-range interactions and manifestations of disorder-related phenomena. However, we find that by radiatively broadening the single meta-molecule resonances and reducing their $Q$-factors can even in the optical regime of metallic resonators lead to intense light confinement and a response that is fundamentally different from that of regular arrays.

\subsection{Asymmetrically split ring resonators}

To determine the effects of disorder, we consider $30 \times 36$ arrays of
ASR meta-molecules \cite{FedotovEtAlPRL2007} arranged in a square
lattice.
We introduce disorder into the meta-molecule positions
by placing each one randomly within a square with side length of a
fraction $D$ of the lattice spacing centered on each unit-cell.
Periodic and disordered  microwave ASR arrays fabricated on a $1.6$mm thick FR4 dielectric substrate were characterized under normal incidence illumination using a pair of linearly polarized broadband horn antennas \cite{papasimakis2009}.

\subsection{Numerical model}

Theoretical analysis of disordered arrays poses a challenge, as, e.g.,  methods exploiting the regular lattice structure~\cite{Kastel07} no longer are possible.
Our numerical model is designed for simulations of a cooperative response~\cite{Lehmberg1970a,Lehmberg1970b,Ishimaru1978,Ruostekoski1997a,JavanainenMFT} in large strongly coupled resonator arrays~\cite{JenkinsLongPRB} that we here
extend to disordered resonator systems. We find that a simplified physical model of a single-mode $RLC$ circuit in a dipolar approximation for each resonator arc provides a sufficient description for the experimentally observed collective radiative properties when the scattering processes between the resonators are incorporated in all orders in the calculations. In large-scale numerical simulations we evaluate the EM response of over 2000 interacting resonators, corresponding to the experimental configurations, and analyze
statistical fluctuations due to disorder by ensemble-averaging over many stochastic realizations..

To numerically model the collective interactions between the meta-molecules in
the metamaterial, consider how each meta-molecule responds to and scatters an
external field.
Currents can flow along each of the arcs, as shown in
Fig.~\ref{fig:schematic}a.  A symmetric current flow in meta-molecule
$\ell$, with an amplitude $d_\ell$, is dominated by
an electric dipole that couples strongly to the incident electric
field.  By contrast, currents in ASR $\ell$ flowing out of phase,
described by an amplitude $m_\ell$,  have a suppressed
electric dipole; they produce radiation into the plane of the array
by the magnetic dipole (and weaker electric quadrupole) \cite{FedotovEtAlPRL2007}.

As explained in App.~\ref{sec:descr-asymm-split}, we model interactions between meta-molecules by decomposing each one
into two resonators, corresponding to the arcs, each of which behaves
like a damped $RLC$ circuit driven by external fields. The dynamics of
arc $j$ ($j=1\ldots 2N$) is described by the oscillator normal
variable $b_j$. The symmetric ($d_\ell$) and antisymmetric ($m_\ell$)
amplitudes are normalised such that
lower arc of unit cell $\ell$ has the amplitude $b_{2\ell-1} = (d_\ell +i m_\ell)/\sqrt{2}$, and the
amplitude of the upper arc $b_{2\ell} = (d_\ell - im_\ell)/\sqrt{2}$.

Current flows produce oscillating multipoles that
couple to the incident field and the fields scattered by other
arcs in the array. The oscillations in each arc are damped at a
decay rate $\Gamma$ which has contributions from an electric dipole
decay rate $\Gamma_e$, a magnetic dipole decay rate $\Gamma_m$, and a
non-radiative damping rate $\Gamma_o$ accounting for Ohmic losses
in the metal and losses in the substrate material.
For the microwave resonators we set $\Gamma_o=0.07\Gamma$ in the numerics
to describe the losses in the substrate.
For the plasmonic resonators we assume that each ASR arc has the Ohmic loss rate given by $\Gamma_o = 0.25 \Gamma$ and $\Gamma_r=\Gamma_m$. The Ohmic loss rate is comparable with those observed for gold rods
Fano resonance experiments~\cite{LiuEtAlNatMat2009} and obtained by Drude-model based estimates~\cite{kuwate}.
To enhance the strength of cooperative interactions, we consider a realistic
array of metallic meta-molecules that are closely spaced with a lattice spacing of $a=0.2\lambda$.

The cooperative response of the metamaterial array emerges from the multiply scattered EM fields between the resonators (App.~\ref{sec:descr-asymm-split}). Each meta-atom is driven by the incident fields and the fields radiated by the other meta-atoms. This leads to the dynamics of the EM response, described by a coupled set of linear equations between the different meta-atoms. The radiative coupling matrix between the arcs yields the collective excitation eigenmodes and the corresponding eigenvalues  (the collective resonance frequencies $\omega_j$ and decay rates $\gamma_j$).

A key element in the EM response of a disordered system is the statistical effect of disorder in the observable quantities. Although our experiment is restricted to a few realizations of the disorder in the resonator positions, in the numerical
simulations we can analyze the statistical properties of the disorder by  stochastically sampling
the positional disorder of the randomly distributed meta-molecules. For each individual stochastic realization of meta-molecule positions, we calculate the scattered fields and the EM response for the quantities of interest. By means of ensemble-averaging over many such realizations, we obtain both ensemble averages and statistical fluctuations of the EM response of the magneto-dielectric array that represent the given statistics of the positional disorder.
Each ASR $\ell$ is located at position $\spvec{r}_\ell = \spvec{R}_\ell + \delta\spvec{r}_\ell$, where $\spvec{R}_\ell$ is the center of the
corresponding unit cell, and $\delta\spvec{r}_\ell$ is the random displacement of the ASR.
As explained in App.~\ref{sec:simul-meas-quant}, we treat the positional displacement as a random variable uniformly distributed within the square interval $x \in (-aD/2, aD/2)$, $y \in (-aD/2, aD/2)$, where $a$ is the periodic array unit cell size and $D$ quantifies the strength of disorder. For a typical observable quantity $O$ of an array of $N$ ASR resonators, we then calculate its averages $\left<O\right>$ and variances $(\Delta O)^2= \left< O^2 \right> - \left<O\right>^2 $ subject to the disorder by ensemble-averaging over a large number of stochastic realizations.

\section{Collective excitation eigenmodes of interacting resonators}

\begin{figure*}
  \centering
  \includegraphics[width=1.8\columnwidth]{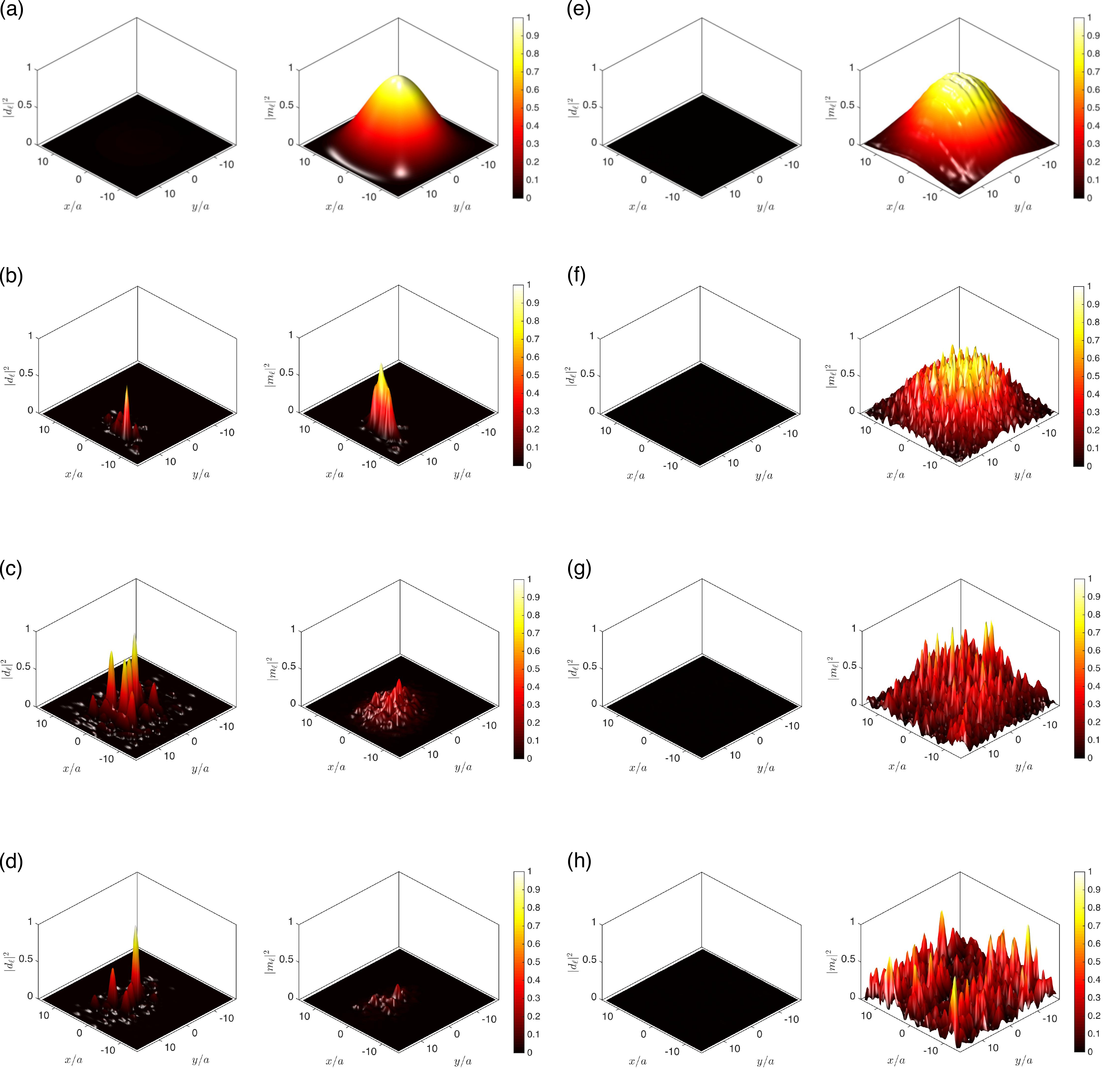}
   \caption{ \textbf{The effects of positional disorder and ASR density on the collective uniform magnetic eigenmode.} (a-d) Electric (left) and magnetic dipole (right) excitations of a single radiative excitation eigenmode
      with a lattice spacing $a=0.28\lambda$ (microwaves). From top to bottom, the ASRs are increasingly displaced from the centers of their respective unit-cells by
    $0\%$ (a), $50\%$ (b), $100\%$ (c) and $150\%$ of the
    corresponding displacements in an experimental sample with $D=0.22$. The corresponding radiative decay rates are $0.21\Gamma$, $0.21\Gamma$, $0.24\Gamma$, $0.28\Gamma$.
    (e-h) The uniform mode in arrays whose ASR
    positions correspond to those on (a-d) scaled by a factor of
    five, giving the regular array (e) a lattice spacing of $a=1.4\lambda$. The corresponding radiative decay rates are $0.91\Gamma$, $0.96\Gamma$, $1.1\Gamma$, $1.2\Gamma$.
  }
  \label{fig:mode_illustration}
\end{figure*}

\begin{figure}
  \centering
  \includegraphics[width=1\columnwidth]{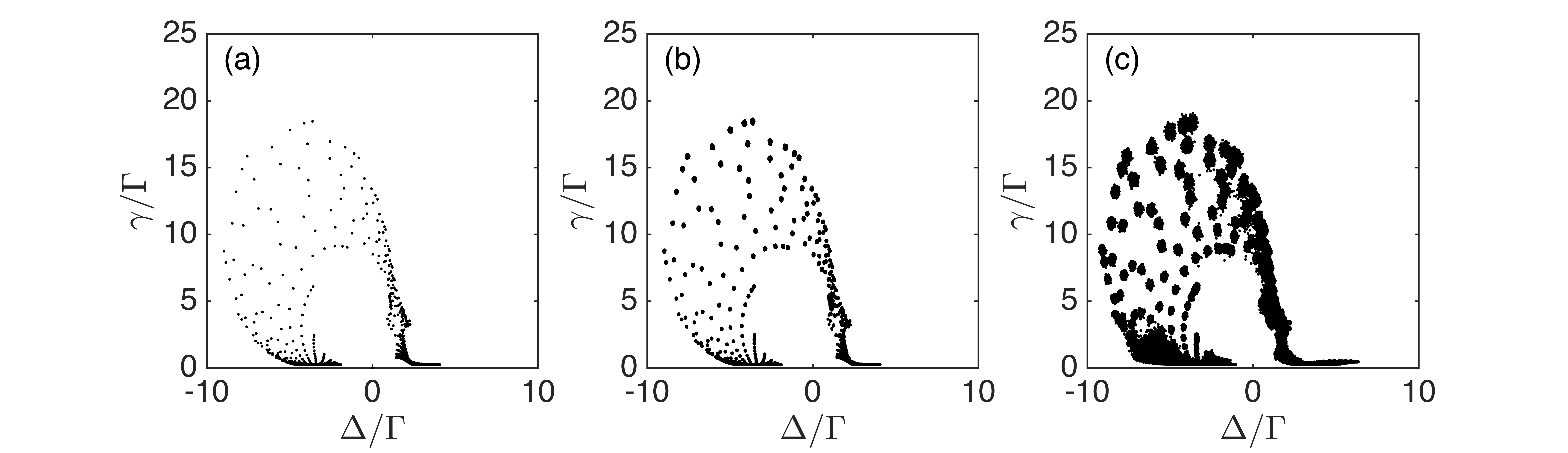}
\caption{\textbf{Collective mode shifts and decay rates of regular and disordered plasmonic ASR arrays.}  Panels (a-c) show the distribution for a regular (a) and two disordered arrays with $D=0.22$ (b) and $D=0.44$ (c), respectively.
  }
  \label{fig:sim_modes_shift_decay_scatter_regular}
\end{figure}

We find that the spatial disorder manifests itself in the collective radiative excitation eigenmodes that are responsible for the transmission resonances of the system.
Even small changes in the positions of the resonators dramatically alter the spatial profiles of the relevant eigenmodes, which in turn has a profound effect on the
EM response of the array.

The emergence of the cooperative effects~\cite{Lehmberg1970a,Lehmberg1970b,Ishimaru1978,Ruostekoski1997a} results from the scattered fields that mediate interactions between resonators.
When the interactions between meta-molecules are strong, as occurs for the subwavelenght lattice spacing of the experimental sample,
the radiative response of a single, isolated meta-molecule is no longer a simple guide to the response of the array; the
metamaterial response becomes a function, not only on properties of individual unit-cells, but on \emph{collective} modes of excitation~\cite{BergmanStroudPRB1980,JenkinsLongPRB}
involving many meta-molecules distributed over the whole of the metamaterial array.
Each mode is characterized by a distinct collective resonance
frequency and decay rate. Here we take the many-body subradiant eigenmode of a regular array of Ref.~\onlinecite{Jenkins2017PRL} in the planar array and show how
this spatially extended mode becomes strongly localized as a function of increasing weak spatial disorder
when we change the resonator positions.

When incident fields drive the regular array at the resonance of the subradiant magnetic eigenmode, the excitation of the electric dipoles across the plane is transferred to a nearly uniform excitation of magnetic dipoles across the entire metamaterial. The transfer of energy between the coherently oscillating electric and magnetic dipoles is possible
because of the asymmetry in the lengths of the arcs within each ASR meta-molecule. About 70\% of this excitation is concentrated on a single eigenmode in which the meta-molecular magnetic dipole amplitudes oscillate in phase with each other, pointing to the direction normal to the plane~\cite{Jenkins2017PRL}; see Fig.~\ref{fig:mode_illustration}. This eigenmode extends over the entire array of over 1000 meta-molecules, has a resonance frequency $\omega_M$ (shifted from that of a single arc in isolation $\omega_0$), and a suppressed subradiant collective radiative decay rate of $0.21 \Gamma$. In Fig.~\ref{fig:mode_illustration} the effect of the array edge can be identified by a reduced excitation amplitude in the outermost unit-cells.

To see how this mode is affected by disorder, we consider one realization of experimental unit-cell displacements $\delta\spvec{r}_{\ell}^{(1)}$, and the collective mode in
arrays whose resonators are partially moved toward those positions. That is, we determine the collective modes for resonators at positions $\spvec{R}_{\ell} + \alpha \delta \spvec{r}_\ell^{(1)}$. We find a dramatic deformation in the profile of the eigenmode even for small values of $\alpha$, as shown in Fig.~\ref{fig:mode_illustration}. For the particular realization of resonator positions, the mode moves to one side of the array and becomes more localized. As the displacement of the resonator positions from the center continues to increase, the mode also gains a stronger contribution from electric dipole excitations. Perhaps surprisingly, the effect of the disorder on the radiative decay rates is
notably weaker, and the subradiant nature of the eigenmode is preserved even for the displacement $D=0.33$ when the radiative decay rate is about $0.28 \Gamma$.

The localization of the mode to a single region of the array is another feature of strong interactions associated with higher densities. In fact, the collective response is very sensitive to the lattice spacing between the resonators owing to the leading $\propto 1/r^3$ contribution to the dipole-dipole interactions. When the resonator positions are scaled so that the underlying regular array has a lattice spacing greater than a wavelength, the response changes drastically. The deformation of the mode profile now consists of multiple regions of excitation distributed over the array. Also the subradiant nature of the mode is almost entirely lost. In the regular array, the radiative decay rate becomes $0.85 \Gamma$ and in the disordered case it quickly reaches the value close to $\Gamma$.

The 30$\times$36 array consists of 1080 unit cells and 2160 collective excitation eigenmodes, collective radiative resonance linewidths and line shifts. All these collective linewidths and line shifts for one specific case with and without disorder are shown as scatter plots in Fig.~\ref{fig:sim_modes_shift_decay_scatter_regular}. The distribution of
collective eigenmode resonance frequencies and decay rates illustrates how many of the collective mode resonance frequencies are shifted from the single-arc resonance. The collective mode decay rates span several orders of magnitude, with the strongly suppressed decay rates of subradiant modes satisfying $\gamma_j\ll \Gamma$.
For small degrees of disorder, there is a class of superradiant modes which look relatively
unaffected by the disorder.  That is, though the disorder shifts collective resonance frequencies of all the modes, in these superradiant modes, this shift is much smaller than the distance to
neighbouring modes.

\section{Far-field detection of back-scattered intensity}

We can directly link the excitation of the giant subradiant eigenmode in Fig.~\ref{fig:mode_illustration}  to the reflected far field.
The excitation patterns in a regular metamaterial array due to an incident plane wave show little visible signs of strong collective interactions
between the resonators. However, under more careful examination the cooperative nature of the metamaterial response manifests itself  in the reflection spectra even in the case of a regular array. In an ASR array interactions can form a Fano resonance in lattices whose individual unit-cells, in isolation, show no such resonance \cite{FedotovEtAlPRL2010,CAIT}.
It was recently shown~\cite{Jenkins2017PRL} that this transmission resonance in a regular array reveals the excitation of the giant many-body subradiant eigenmode of Fig.~\ref{fig:mode_illustration}.
In an array where the meta-molecules do not interact the anti-symmetric oscillations of a meta-molecule are
damped as strongly as symmetric oscillations, and the array exhibits no transmission resonance.

\begin{figure*}
  \centering
  \includegraphics[width=2\columnwidth]{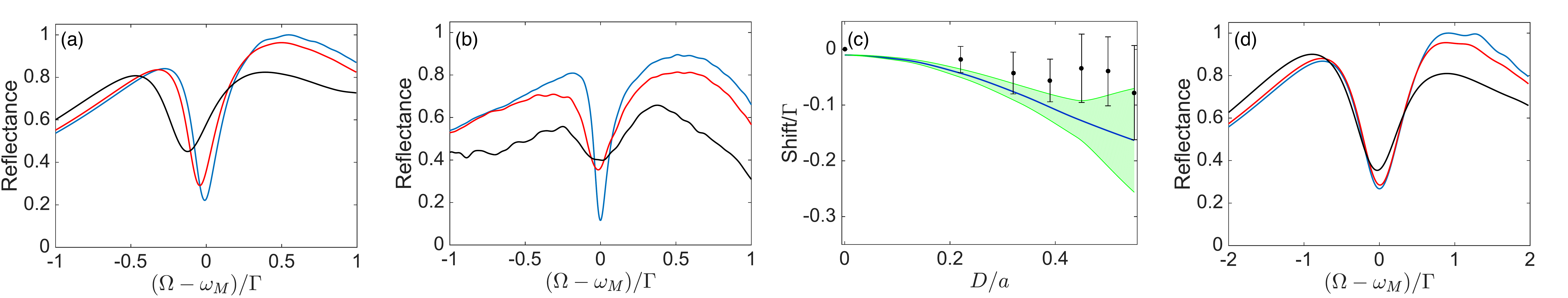}
  \caption{\textbf{Far field spectrum of microwave and optical plasmonic arrays
  with varying degree of disorder}. The
    back-scattered intensity from the microwave array \textbf{a.} as calculated by the theoretical model
    and  \textbf{b.} measured in the experiments. The blue lines represent
    scattering from ordered arrays ($D=0$), and the red (black) lines
    represent scattering from arrays with disorder parameter $D=0.22a$
    ($D=0.44a$).
    \textbf{c.}
    the experimentally measured and theoretically calculated shift of the transmission resonance as a function of disorder (corresponding to \textbf{a-b}). The filled
    regions indicate values within one standard deviation of the
    ensemble average in theoretical simulations and the markers indicate averages obtained
    experimentally from five realizations of resonator positions, with
    the error bars spanning one standard deviation.
    \textbf{d.} The back-scattered intensity from the optical plasmonic array
    with a lattice spacing of
    $0.2\lambda$ and an Ohmic loss rate of $25\%$ of the total loss
    rate as calculated by the theoretical model. The degrees of disorder being shown are
    $D = 0, 0.22,$ and $0.44$. All theoretical simulations are ensemble
    averaged over 1024 stochastic realizations.
  }
  \label{fig:far_field_exp_th_var_disord}
\end{figure*}

Figure~\ref{fig:far_field_exp_th_var_disord} shows how the collective transmission resonance of the uniform response of a regular lattice
still persists even in the presence of disorder. The disorder alters the interactions between the resonators and the quality of the resonance (determined by contrast and width) is reduced.
In  Fig.~\ref{fig:far_field_exp_th_var_disord}a,b we show a side-by-side comparison between experimental observations and large-scale numerical simulations of the back-scattered intensity
(see Apps.~\ref{sec:descr-asymm-split} and~\ref{app:exp}) from disordered microwave metamaterial arrays.
A narrow dip occurs in the back-scattered intensity at a frequency
around $11$GHz.
Our numerical model demonstrates that this resonance arises solely as a
result of interactions between meta-molecules.
The different curves indicate how starting from a regular lattice and then
increasing
the degree of positional disorder considerably affects the nature of the
resonance. The contrast, width, and shift of the resonance vary from sample
to sample, and in our numerical simulations we found that this variance increases with the degree of positional
disorder in the array.  This variation is a manifestation of the
dependence of interactions on the relative positions of the
resonators, and shows how the fluctuations of the positions are mapped onto the fluctuations of the transmission and reflection resonance properties of the fields.

Both experimental measurements and
theoretical calculations, show that (Fig.~\ref{fig:far_field_exp_th_var_disord}a,b), on average, the contrast of the transmission
resonance is reduced and the resonance becomes broader as the resonators
take on more random positions relative to one another. While the theoretical model qualitatively captures the essential features of the positional  disorder,
the spectrum for the experimental system is, however, more sensitive
to disorder than that of the theoretical model.  This could
be because of the interaction of the meta-molecules with the substrate that our
model does not account for.  Additionally, because our simplified
model takes each arc to be a point dipole, it may underestimate
the interaction between arcs as their outer edges move closer
together. The disorder of $D=0.22$ in the experimental system
therefore corresponds to an effective disorder of $D=0.44$ in the
model array.

Measuring the shifts of resonances in strongly coupled resonant emitter systems has in recent years attracted considerable attention. Such collective resonance shifts are frequently also referred to as ``cooperative Lamb shifts''~\cite{Friedberg1973}, and have been detected in systems composed, e.g., of nuclei~\cite{ROH10}, ions~\cite{MeirLamb}, thermal atoms confined inside vapor cells~\cite{Keaveney2012}, and cold trapped atomic ensembles~\cite{Jennewein_trans,Roof16,Ye2016,Dalibard_slab}.
The origin of the shifts is the resonant interaction due to the surrounding emitters that modifies the resonance linewidths and line shifts and therefore results in an effective renormalization of the resonance frequency.
The observed shifts have been found to differ from the basic Lorentz-Lorenz shift~\cite{Jackson,BOR99}, due to collective interactions.
In dense and homogeneously broadened ensembles, resonance shifts can notably differ from those of inhomogeneously-broadened ensembles, and can violate standard textbook results of
electrodynamics~\cite{Javanainen2014a,JavanainenMFT,Jenkins_thermshift,Jennewein_trans,Dalibard_slab}.
Here, we observe a shifting of the resonance toward the red as disorder is increased,
which provides a clear indication of the collective nature of the EM response of the system.
This pattern is borne out both in the
theoretical model and in the experiment, as shown in
Fig.~\ref{fig:far_field_exp_th_var_disord}c. The statistical fluctuations of the shift are
calculated by considering 1024 distinct realizations of resonator positions within each unit-cell (see App.~\ref{sec:simul-meas-quant}).

By means of comparing the collective eigenmodes of interacting resonators with the far-field spectrum, we have established how the scattered fields convey information about the positional disorder and  collective effects in a metamaterial. Strong Ohmic losses in optical and near-infrared plasmonic metamaterials, however, can curtail the role of interactions in
the material's response, since they limit the number of times a photon can scatter before it is absorbed~\cite{JenkinsLongPRB}.
Nonetheless, when plasmonic meta-molecules are closely spaced, we still find signatures of a strong collective response.
The magnetic mode responsible for the transmission resonance in
the microwave array, can no longer be employed to produce as high a
quality transmission resonance.
The material still supports a transmission resonance, though be it a
broader one, as
shown in Fig.~\ref{fig:far_field_exp_th_var_disord}d.
However -- despite the reduced collective effects due to stronger absorption --
the introduction of disorder in the resonator positions even in a plasmonic array  notably disrupts the
transmission resonance (Fig.~\ref{fig:far_field_exp_th_var_disord}d).

We would like to note that although fabrication imperfections are more prominent in metamaterias for the optical part of the spectrum, the degree of disorder required for applications is substantially higher than the fabrication precision, and hence the realization of such disordered metamaterials will not be significantly hindered by manufacturing limitations.

The reduction of the transmission resonance with increased
disorder can be understood by considering the excitations within each
meta-molecule.
In a regular array, the transmission resonance occurs when the
anti-symmetric oscillations
are excited at the expense of the symmetric excitations
\cite{JenkinsLineWidthNJP,FedotovEtAlPRL2010}.
Since the electric dipoles produced by the symmetric
meta-molecule excitations radiate in the forward and backward directions,
and the anti-symmetric excitations do not, a lack of symmetric
meta-molecule excitations implies a lack of reflectance.

\section{Concluding remarks}

We have analyzed the microscopic principles of macroscopic EM response in disordered metamaterials. Our work indicates that strongly interacting metamaterials, where the discrete nature of the resonators and the field-induced correlations become relevant, could be an especially fruitful avenue in the design of novel metamaterial functionalities and in exploitation of the effects of positional disorder. Planar arrays of ASRs provide a model example of a system where strong interactions can be achieved. In such strongly-interacting systems, cooperative response, that is absent in weakly coupled ordinary resonator arrays and `hidden' in the response of strongly coupled resonators in regular arrays, manifests itself in the presence of disorder. The non-uniform subwavelength-scale response and localization due to disorder indicate the breakdown of effective continuous-medium theories for electrodynamics that assume each resonator interacting with the average behavior of all the other surrounding resonators, while the system is composed of discrete emitters. This is analogous to the breakdown of mean-field theories in condensed-matter physics where enhanced interparticle interactions lead to correlated system response.

Our microscopic analysis of the electrodynamics of disordered metamaterials paves a way for novel design paradigms in artificial EM materials that are based on the discrete nature of the metamaterial lattice and the resulting cooperative effects rather than effective medium considerations. Our approach allows to tailor the metamaterial functionalities both at the microscopic (meta-molecule) as well as the macroscopic (array) scale. The localization of the array collective modes in combination with the prescribed multipole character of the excitation at the resonator level holds promise for a number of applications, including control of emitter rate and directivity, sensing, nonlinear optics, focusing.

Light transmission in disordered media of resonant scatterers generally attracts broad interest in many different physical systems. These include natural media, formed by atoms, as well as those composed of artificial atoms. Although, for example, Anderson localization of light in 1D is well established and analyzed (as an example of a recent experiment, see Ref.~\onlinecite{Sapienza17}), there is still considerable debate whether light can even undergo Anderson localization in 3D systems of dipolar scatterers due to disorder~\cite{Skipetrov14,MaretNJP}.
Moreover, the framework for the analysis of disordered metamaterial arrays presented here can be readily extended to include higher order terms of the multipole expansion beyond the electric and magnetic dipoles \cite{Watson2016,Watson2017}. Finally, engineering collective modes and utilizing disorder will be of particular interest in the studies of highly nonlinear superconducting metamaterials that display, e.g., resonator synchronization \cite{Lazarides2013,Ustinov2015,Anlagepre17,Fistul2017}.

\begin{acknowledgments}
We acknowledge financial support from the EPSRC (EP/G060363/1, EP/M008797/1, EP/P026133/1), the Leverhulme Trust, the Royal Society, and the MOE Singapore Grant No. MOE2011-T3-1-005. We also acknowledge the use of the IRIDIS High Performance Computing Facility at the University of
Southampton.
\end{acknowledgments}

\appendix

\setcounter{equation}{0}
\setcounter{figure}{0}
\renewcommand{\theequation}{A\arabic{equation}}
\renewcommand{\thefigure}{A\arabic{figure}}

\section{Asymmetrically split rings and the model for collective interactions}
\label{sec:descr-asymm-split}

\subsection{Asymmetrically split ring meta-molecule}

The general formalism to describe interacting magnetodielectric resonators is presented in \cite{JenkinsLongPRB}.
We model each unit-cell resonator, labelled by index $\ell = 1 \ldots 30\times 36$ as an ASR meta-molecule supporting two types of current oscillation: (i) currents flowing
symmetrically and (ii) with currents flowing anti-symmetrically.  The amplitude and phase of these oscillations within meta-molecule $\ell$ are described by the complex amplitudes $d_\ell$ and $m_\ell$ respectively. Symmetric oscillations possess a net electric dipole proportional to $d_\ell \unitvec{d}$, while owing to the curvature, anti-symmetric oscillations are dominated by a
magnetic dipole proportional to $m_\ell \unitvec{m}$ with a small electric quadrupole.
Each ASR $\ell$ is located at position $\spvec{r}_\ell = \spvec{R}_\ell + \delta\spvec{r}_\ell$, where $\spvec{R}_\ell$ is the center of the
corresponding unit cell, and $\delta\spvec{r}_\ell$ is the random displacement of the ASR. The amplitudes for symmetric and antisymmetric
oscillations $d_\ell$ and $m_\ell$ are normalized such that the total
energy contained in an ASR excitation is proportional to $|d_\ell|^2 +
|m_\ell|^2$. Each unit-cell resonator is decomposed
into two asymmetric arcs, or meta-atoms, each of which behaves like a single-mode damped $RLC$ circuit with resonance frequency $\omega_j$ driven by external fields.
If the split rings were symmetric, the individual meta-atoms would have identical resonance frequencies $\omega_j = \omega_0$.
An asymmetry in the arc lengths shifts the meta-atom resonance frequencies by $\delta\omega$ so that for ASR $\ell$
\begin{subequations}
  \label{eq:meta_at_freqs}
  \begin{eqnarray}
    \omega_{2\ell-1} = \omega_0 - \delta\omega \textrm{ ,}\\
    \omega_{2\ell} = \omega_0 + \delta\omega \textrm{ .}
  \end{eqnarray}
\end{subequations}
The current excitations in the meta-atoms interact with the propagating field and radiate electric and magnetic fields. Each meta-atom is treated in the point dipole approximation. With this approach the EM properties of the meta-molecules can be obtained by assigning each arc an electric dipole $\spvec{d}_j(t) = d_j(t)\unitvec{e}_y$  and
magnetic dipole $\spvec{m}_j(t) = m_j(t)\unitvec{m}_j$, where $\unitvec{m}_{2\ell} = -\unitvec{m}_{2\ell-1} \equiv \unitvec{m} = \unitvec{e}_z$.

The oscillating current excitations in each meta-atom $j$ ($j=1\ldots 2N$) are described by the oscillator normal mode amplitude $b_j$, and the
lower arc of unit cell $\ell$ has the amplitude $b_{2\ell-1} = (d_\ell +i m_\ell)/\sqrt{2}$, and the
amplitude of the upper arc $b_{2\ell} = (d_\ell - im_\ell)/\sqrt{2}$.
Here, and in the rest of the discussion, we assume that all the field and resonator amplitudes refer to the slowly-varying versions of the
positive frequency components of the corresponding variables,
where the rapid oscillations $e^{-i\Omega t}$ ($k=\Omega/c$) due to the frequency, $\Omega$, of the incident wave have been factored out in the rotating wave approximation.
Oscillations in each arc are damped (at rate $\Gamma$) by electric dipole radiation,
magnetic dipole, radiation and ohmic losses at rates $\Gamma_e$,
$\Gamma_m$ and $\Gamma_o$, respectively.
The strength of radiative interactions between
dipoles of the arcs of each meta-molecule are governed by their
separation $u$.
The upper and lower arcs are located at $\spvec{r}_j +
(u/2)\unitvec{e}_y$ and $\spvec{r}_j - (u/2)\unitvec{e}_y$,
respectively.

\subsection{Collective EM-mediated interactions}

The scattered fields from the arcs are given by ${\bf E}_{S}=\sum_j {\bf E}_{S}^{(j)}$ and ${\bf H}_{S}=\sum_j {\bf H}_{S}^{(j)}$ where the contributions from the meta-atom $j$ read
\commentout{
\begin{align}
{\bf E}_{\text{sc},j}({\bf r},t)&=\frac{k^3}{4\pi\epsilon_0}\sum_j
\bigg[{\sf G}({\bf r} -{\bf r}_j){\bf d}_j(t)\nonumber\\
&+\frac{1}{c}{\sf G}_\times({\bf r} - {\bf r}_j)
{\bf m}_j(t)\bigg],\label{eq:Esc}\\
{\bf B}_{\text{sc},j}({\bf r},t)&=\frac{\mu_0k^3}{4\pi}\sum_j
\bigg[{\sf G}({\bf r} -{\bf r}_j){\bf m}_j(t)\nonumber\\
&-c{\sf G}_\times({\bf r} - {\bf r}_j){\bf d}_j(t)\bigg],
\label{eq:Bsc}
\end{align}
}
\begin{align}
{\bf E}_{S}^{(j)}({\bf r},t)&=\frac{k^3}{4\pi\epsilon_0}
\bigg[{\sf G}({\bf r} -{\bf r}_j){\bf d}_j
+\frac{1}{c}{\sf G}_\times({\bf r} - {\bf r}_j)
{\bf m}_j\bigg],\label{eq:Esc}\\
{\bf H}_{S}^{(j)}({\bf r},t)&=\frac{k^3}{4\pi}
\bigg[{\sf G}({\bf r} -{\bf r}_j){\bf m}_j
-c{\sf G}_\times({\bf r} - {\bf r}_j){\bf d}_j\bigg]\,.
\label{eq:Bsc}
\end{align}
The dipole radiation kernel ${\sf G}({\bf r})$ determines
the electric (magnetic) field at ${\bf r}$, from an oscillating electric (magnetic) dipole
at the origin~\cite{Jackson}.  For a dipole with an amplitude $\hat{\bf d}$, the expression reads
\begin{align}
{\sf G}&({\bf r})\,\hat{\bf d}=
(\hat{\bf n}\!\times\!\hat{\bf d}
)\!\times\!\hat{\bf n}{e^{ikr}\over kr}\nonumber\\
& +[3\hat{\bf n}(\hat{\bf n}\cdot\hat{\bf d})-\hat{\bf d}]
\bigl[ {1\over (kr)^3} - {i\over (kr)^2}\bigr]e^{ikr}
-{4\pi \hat{\bf d}\,\delta(k {\bf r})\over3}\,,
\label{eq:DOL}
\end{align}
where $\hat{\bf n} = {{\bf r}/ r}$.  The contact term is included to satisfy the Gauss law, and we interpret Eq.~\eqref{eq:DOL} in
such a way that the integral over an infinitesimal volume enclosing the origin
of the other terms vanishes.

The cross kernel ${\bf G}_\times({\bf r})$ describes the electric (magnetic) field at ${\bf r}$ of an oscillating magnetic (electric) dipole
at the origin. For a dipole with an amplitude $\hat{\bf d}$, we have
\beq
{\sf G}_\times ({\bf r})\,\hat{\bf d}= {i\over k} \nabla\times {e^{ikr}\over kr}\, \hat{\bf d}\,.
\eeq

Each meta-atom is driven by the incident fields, $\cbE_0(\spvec{r},t)$ and $\cbH_0(\spvec{r},t)$,  and the fields scattered by all the other resonators in the system,
\begin{align}
{\bf E}_{\text{ext}}({\bf r}_j,t)& = \cbE_0(\spvec{r},t)
+ \sum_{l\ne j}{\bf E}_{S}^{(l)}({\bf r},t),
\label{eq:Eext}\\
{\bf H}_{\text{ext}}({\bf r}_j,t)& = \cbH_0(\spvec{r},t) +
\sum_{l\ne j}{\bf H}_{S}^{(l)}({\bf r},t)\,,
\label{eq:Bext}
\end{align}
where the scattered fields are given by Eqs.~\eqref{eq:Esc} and~\eqref{eq:Bsc}.

Owing to the coupling between the current oscillations and the scattered EM fields we obtain the coupled dynamics for the arc variables $\colvec{b} \equiv (b_1, b_2, \ldots, b_{2N})^T$~\cite{JenkinsLongPRB},
\beq
\dot{\colvec{b}} = \mathcal{C}\colvec{b} + \colvec{F}(t)\,.
\label{eqnmotion}
\eeq
In the dipole approximation, the normal mode amplitudes (unnormalized) of each meta-atom are linked with its electric and magnetic dipoles
\beq
 b_j(t) =
 \sqrt{\frac{k^3}{12\pi\epsilon_0}}
 \Bigg[
\frac{d_j}{\sqrt{\Gamma_{e}}}
  +
 i \frac{m_{j}}{c\sqrt{\Gamma_{m}}}
 \Bigg]
 \,\text{.}
 \label{eq:b_multipole}
 \eeq
The matrix $\mathcal{C}$ accounts for radiative and non-radiative decay, the asymmetry of the individual unit-cells, and crucially, the EM interactions between the unit-cells
mediated by the scattered field (that incorporate the retardation effects with short- and long-range interactions). The vector $\colvec{F}$ represents the driving of the current in each arc caused by the incident field and can be expanded to the eigenmodes of the array with amplitudes $f_n$.

Explicitly,  the coupling matrix is
\begin{align}
  \mathcal{C} =& -\frac{\Gamma_e + \Gamma_m}{2}\mathbf{1} +i\frac{3}{4}
  \left(\Gamma_e \mathcal{G}_e + \Gamma_m\mathcal{G}_m \right)
  \nonumber\\
  & +
  \frac{3}{4} \sqrt{\Gamma_e \Gamma_m}\left(\mathcal{G}_\times +
    \mathcal{G}_\times^T\right) \text{,}
    \label{eq:CopulingMatC}
\end{align}
and the driving field contribution is given by
\begin{equation}
  \label{eq:ForcingFunc}
  \colvec{F} = i
  \frac{\mathcal{E}_{in}(t)}{\sqrt{2\omega_0 L}} \text{,}
\end{equation}
where $\colvec{\mathcal{E}}_{in}$ is the electromotive force
induced by the driving field.
The dimensionless coupling matrices $\mathcal{G}_e$, and
$\mathcal{G}_m$ result from interactions with the  electric or
magnetic fields scattered from the electric or magnetic dipoles respectively,
while the matrix $\mathcal{G}_\times$ accounts for the electric (magnetic)
fields produced by the magnetic (electric) dipoles.
Equation (\ref{eqnmotion}) corresponds to the integral representation of Maxwell's
wave equations, and can be efficiently solved
as a linear system.

In a metamaterial array, we have a system of $N$ ASR meta-molecules, or $2N$ single-mode resonator arcs.
These possess $2N$ collective eigenmodes of current oscillation, with corresponding collective resonance frequencies and decay rates.
Each collective eigenmode corresponds to an eigenvector of the matrix $\mathcal{C}$ with the eigenvalues given by
\begin{equation}
\lambda_j= -\frac{\gamma_j}{2}-i \delta\omega_j\,,
\label{eq:eigenvalue}
\end{equation}
where the decay rate is $\gamma_j$ and the shift of the resonance frequency with respect to the arc frequency $\omega_0$ is given by $\delta\omega_j$.
In this work we calculate the eigenmodes of the 30$\times$36 array of 2160 meta-atoms.

The simulation techniques described here are quite general and can be adjusted to the studies of cooperative phenomena in other point scatterer systems~\cite{Lagendijk,devries98}, including those in atomic ensembles~\cite{Jenkins2012a,Javanainen2014a}. Recently, similar simulation techniques based on the point-dipole scatterers have been applied in the design and modelling of metasurfaces \cite{Drsmith2017,Drsmith2017b}.

\section{Simulating ensemble averages of disordered system using stochastic sampling}
\label{sec:simul-meas-quant}

Random displacements in meta-molecule positions profoundly affect the EM response of the metamaterial.  This
happens because a single arc is driven, not only by the incident field, but by the fields emitted by all other arcs in the array.
The influence of those scattered fields depend sensitively on the arcs' relative positions, particularly when the sample is dense.
Every observable quantity is therefore a function of the displacements of all of the ASRs in the array.
Numerical simulations allow the calculation of the statistical properties of the EM response in the presence of positional disorder
through stochastic sampling.

Formally, our sampling technique is described as follows.  Consider some observable quantity $O$ of an array of $N$ ASR resonators.  For a
specific realization of displacements, ASRs are displaced from the centers of their respective lattice sites $\spvec{r}^{(\mathrm{lat})}_\ell$
by $\delta\spvec{X}_\ell$. The observable quantity could be, for example, the magnetic dipole intensity $|m_\ell|^2$ of a particular
resonator $\ell$, the maximum ASR excitation taken over all elements of the array (excluding the ten outer most unit cells), or
the back scattered intensity from the array in the far-field.  The key feature of the simulated observable is that a single realization
of ASR positions maps directly to a specific observed value.  So, the average value of the observable is
  \begin{equation}
    \label{eq:StochEq1}
    \left\langle O \right\rangle = \int d^3\delta x_1\ldots d^3\delta
    x_N\, O(\spvec{r}_1,\ldots,\spvec{r}_N)
    P(\delta\spvec{x}_1,\ldots,\delta\spvec{x}_N)
  \end{equation}
where $P(\delta\spvec{x}_1,\ldots,\delta\spvec{x}_N)$ is the joint
probability distribution for displacements of ASRs from the centers of
their lattice sites, and $\spvec{r}_\ell \equiv
\spvec{r}^{(\mathrm{lat})}_\ell + \delta\spvec{x}_\ell$.
In an array with a degree of disorder $D$, we take the displacements
$\delta\spvec{X}_\ell$ to be independent identically distributed
random variables with a uniform distribution within a square of
side length $Da$ centered on the origin of the $xy$ plane, where $a$ is
the lattice spacing of the unperturbed regular array.

We calculate the average quantities by sampling $\mathcal{N}$
realizations of ASR displacements
$\delta\spvec{X}_1^{(n)}, \ldots, \delta\spvec{X}_\ell^{(n)}$
($n=1,\ldots,\mathcal{N}$, $\ell = 1,\ldots,N$) from the joint probability distribution
$P$, and averaging the desired observable over all realizations.  This yields
\begin{equation}
  \label{eq:mc_average}
  \left\langle O \right\rangle \approx \frac{1}{\mathcal{N}}
  \sum_{n=1}^{\mathcal{N}}
  O(\spvec{X}_1^{(n)},\ldots,\spvec{X}_N^{(n)}) \,\textrm{,}
\end{equation}
where $\spvec{X}_\ell^{(n)} \equiv \spvec{r}^{(\mathrm{lat})}_\ell +
\delta\spvec{X}_{\ell}^{(n)}$ is the position of ASR $\ell$ in
realization $n$ of ASR positions. The statistical variances are calculated analogously
 \begin{equation}
  \label{eq:mc_vari}
(\Delta O)^2= \left< O^2 \right> - \left<O\right>^2  \,.
\end{equation}

\section{Resonance contrast and linewidth estimation}
For each realization of meta-atom positions, we determine the frequency $\delta$ at which the reflected intensity is at a minimum $I_{\mathrm{min}}$. This is the frequency of the transmission resonance for this specific realization. We then determine the frequencies at which the reflected intensity reaches its maxima on either side of the resonance. We define the contrast with respect to the lesser of these two intensities $\tilde{I}$. The contrast of the resonance is then $1-I_{\mathrm{min}}/\tilde{I}$ so that the resonace has unity contrast if suppression of the back-scattered field were perfect, and the contrast is zero when there
is no transmission resonance at all. The spectral width of the resonance is the difference between the smallest frequency above $\delta$ and the greatest frequency below $\delta$ for which the reflected intensity takes the value $(I_{\mathrm{min}} + \tilde{I})/2$.

\section{Experimental methods}
\label{app:exp}
\subsection{Samples}
The experimentally studied samples consisted of regular and disordered ASR arrays fabricated by etching a 35 $\mu$m copper cladding on a FR4 PCB substrate of 1.6 mm thickness. Each ASR has an inner and outer radius of 2.8 and 3.2 mm, respectively. The metamolecules were arranged in a $30\times 36$ square lattice with
lattice spacing $a = 7.5$ mm. Disorder was introduced by displacing the center of each meta-molecule
according to a random uniform distribution defined over the square interval $x \in (-aD/2, aD/2)$, $y \in (-aD/2, aD/2)$, where $D$ is the degree of disorder. We consider metamaterial arrays with different degrees of disorder ($D=0.22, 0.32, 0.39, 0.45, 0.5, 0.55$). For each degree of disorder, we have constructed five samples with different realizations of unit-cell positions.

\subsection{Far-field measurements}
The far-field response of regular and disordered microwave ASR arrays was characterized in an anechoic chamber using a pair of linearly polarized horn antennas (Schwarzbeck BBHA 9120D) and a vector network analyser (Agilent E8364B). The strength of the backscattered radiation from the metamaterial arrays was characterized by measuring their reflectivity under normal incidence illumination. The polarization of the incident and the detected reflected wave was fixed along the arcs of the ASRs.

\end{document}